\documentclass{article}
\usepackage{ijcai18}

\usepackage{times}
\usepackage{xcolor}
\usepackage{soul}
\usepackage[utf8]{inputenc}
\usepackage[small]{caption}
\usepackage{graphicx}
\usepackage{caption}
\usepackage{subcaption}
\usepackage{booktabs}
\usepackage{blindtext}
\usepackage{amsmath}
\usepackage{multirow}
\usepackage{color}
\usepackage{makecell}

\title{Automatically Infer Human Traits and Behavior from Social Media Data}

\author{Shimei Pan\\ 
Department of Information Systems \\
University of Maryland, Baltimore County \\
shimei@umbc.edu
\And 
Tao Ding\\
Department of Information Systems \\
University of Maryland, Baltimore County \\
taoding01@umbc.edu
}

\begin{document}

\maketitle

\begin{abstract}
Given the complexity of human minds and their behavioral flexibility, it requires sophisticated data analysis to sift through a large amount of human behavioral evidence to model human minds and to predict human behavior. People currently spend a significant amount of time on social media such as Twitter and Facebook. Thus many aspects of their lives and behaviors have been digitally captured and continuously archived on these platforms. This makes social media a great source of large, rich and diverse human behavioral evidence. In this paper, we survey the recent work on applying machine learning to infer human traits and behavior from social media data. We will also point out several future research directions. 

\end{abstract}
\section{Introduction}
People currently spend a significant amount of time on social media to express opinions, interact with friends and families, share ideas and thoughts, provide status updates and organize/participate events and activities. According to Nielsen’s 2016 Social Media report, on average, Gen X (ages 35-49) spent 6 hours and 58 minutes and Millennials (ages 18-34) spent 6 hours and 19 minutes per week on social media in the US.  As a result, many aspects of their lives have been digitally captured and continuously archived on social media. 

With recent advent of big data analytics and the availability of a large amount of user-generated content, data-driven human trait and behavioral analysis has increasingly been used to better understand human minds and predict human behavior. Prior research has demonstrated that by analyzing the information in a user’s social media account, we can infer many latent user attributes such as political leaning~\cite{pennacchiotti2011machine,o2013using,kosinski2013private,benton2016learning}, brand preferences~\cite{pennacchiotti2011machine,yang2015using}, emotions~\cite{kosinski2013private}, mental disorders~\cite{de2013predicting,vedula2017emotional}, personality~\cite{kosinski2013private,schwartz2013personality,liu2016analyzing}, substance use~\cite{kosinski2013private,ding2017multi} and sexual orientation~\cite{kosinski2013private}. In addition, understanding individual traits and behavior has numerous real life applications including public health~\cite{giota2013role}, marketing \cite{yang2015using,ding2016personalized} and politics \cite{chirumbolo2010personality}.


We believe user generated data on social media is ideal for data-driven user trait and behavior analysis due to its unique characteristics: (1) large scale: it includes behaviors of a large number of social media users (e.g., millions users); (2) comprehensive: it contains a large number of behavioral markers from diverse sources (e.g., text posts, image posts, likes, and friendship); (3) longitudinal: it follows behaviors of social media users continuously over a long period of time (e.g., years); (4) objective: the analysis is based on the natural behavioral data automatically, continuously and objectively collected in an open environment.

In this survey, we summarize the recent advances in automated human traits and behavior inference from social media data. We included 24 studies published as full-length papers at one of the top AI, data mining, social media and multi-diciplinary behavior conferences and journals in the last ten years. We focus on papers that employ machine learning techniques to automatically infer or predict latent attributes or behaviors of individuals from public social media data.   


\section{Overview of Studies}
\begin{table*}[ht!]
\small
\centering
    \caption{Summary of Studies}
    \label{tab:overview}
    \begin{tabular}{@{}p{4.1cm}p{3.5cm}p{4.5cm}p{1cm}p{3cm}@{}}
      \toprule
       \textbf{Paper} & \textbf{Platform} &\textbf{Source Data Type} & \textbf{Size}  & \textbf{Predicted Target}  \\
       \midrule
       \cite{pennacchiotti2011machine} & Twitter& tweet, user profile & 10,338 & political leaning \\
                                                                       &&& 6,000 &  ethnicity\\
                                                                       &&& 10,000 & user preference \\\hline 
       \cite{yang2012we} & Twitter& tweet &  & hashtag adoption\\
       									&&&1,029& (Politic) \\
                                        &&&	8,029 & (Movie)\\
                                        &&& 15,038 & (Random)\\\hline 
        \cite{de2013predicting} & Twitter &post, social network, user profile &476& depression \\\hline 
       \cite{zhang2013predicting} & Facebook& like & 13,619 & purchase behavior \\\hline
       \cite{o2013using} & Twitter & tweet, social network & &political leaning \\
       										 &&&4,200& (Election day)\\
                                             &&&1,643&  (Tweetcast)\\\hline 
       \cite{kosinski2013private} &Facebook&likes&58,000&age, gender,personality\\
                                                 &&& 46,027 &relationship\\
                                                 &&& 1,200 &substance use\\
                                                 &&& 18,833& religion \\
                                                 &&&  9,752&political leaning\\\hline
       \cite{schwartz2013personality} &Facebook&post & 75,000& age,gender,personality \\\hline
       \cite{chen2014understanding} & Reddit & post & 799 & human value \\\hline
       
      \cite{gao2014modeling} &Twitter&tweet, retweet&&attitude\\
      &&&5,387& (Topic:Fracking)\\
      &&&2,593& (Topic:Vaccination)\\\hline
       \cite{youyou2015computer} &Facebook& like &70,520 & personality \\\hline
       \cite{yang2015using} & Twitter & tweet &608&brand preference\\\hline
       \cite{preoctiuc2015analysis} & Twitter & tweet,user profile & 5,191 &  occupation\\\hline
       \cite{lee2015will}&Twitter&user profile, tweet, social network& &retweet action\\
       &&&1,902& (Public safety)\\
       &&&1,859& (Bird flu)\\\hline
       \cite{shen2015study} &Facebook&like,tweet&1,327&personality\\\hline
       \cite{bhargava2015unsupervised} &Facebook&post&488&interest\\\hline
       \cite{song2015interest} & Twitter,Facebook,Quora & post, bio description & 1,607 & interest\\\hline 
       \cite{hu2016language}&Twitter&tweet&9,800&occupation\\\hline
       \cite{song2016volunteerism} &Twitter,Facebook,Linkedin&tweet, social network &5,436 &volunteerism\\\hline
       \cite{liu2016analyzing} & Twitter& profile picture & 66,000 & personality \\\hline
       \cite{benton2016learning}&Twitter&tweet,social network& 102,328 & topic engagement \\
       												&&&500&friend recommendation\\
                                                    &&&383 & age, gender \\
                                                    &&&396 & political leaning\\\hline 
       \cite{ding2017multi} & Facebook & status, like & 1,200 & substance use\\ \hline
       \cite{vedula2017emotional} & Twitter & tweet, social network & 3215 & depression\\\hline
         \cite{preoctiuc2017beyond} &Twitter& tweet& 3,938& political leaning \\\hline
         \cite{singh2017toward}&Instagram & caption, comments, image&699& cyberbullying\\
      \bottomrule
    \end{tabular}
\end{table*} 
Table~\ref{tab:overview} lists the 24 papers included in our survey.  For each paper, we summarize information related to the dataset such as the ``Platform" from which the social media data were collected (e.g., Twitter, Facebook, Reddit, Instagram and Quora) and ``Size", which is the number of people with ground truth labels in the dataset. In addition, we also lists ``Source Data Type", to indicate the different types of social media data used in the study. Here, 
\textit{text} refers to user-generated text data that include user posts (e.g., tweets or status update on Facebook), profile description, and comments; 
\textit{like} refers to things a social media user likes such as books, people, music, movies, TV shows, photos and products; 
\textit{user profile} includes demographic information (gender, age occupation, relationship status etc.) and aggregated statistics of a user account (the number of friends, followers, followings etc.); 
\textit{posting activity} which includes a set of statistics describing a user's posting behavior on social media such as the total number of tweets/retweets, the average number of posts per day, the number of likes/replies for each post; 
\textit{image} which includes the profile and background photos and also the iamge posts shared on social media; 
\textit{social network} which refers to social connections between different user accounts such as friendship networks on Facebook and follower/following relations on Twitter. 

Although social media provide us an opportunity to easily track a large number of heterogeneous user behavioral data, the characteristics of social media data also bring significant challenges to data analysis. For example, the text and images are unstructured data. Making sense out of unstructured data is alway a big challenge. User likes is very sparse and high dimensional. For example,  the likes data in ~\cite{ding2017multi} includes ~10 millions unique like dimensions. It is also not easy to search and analyze a large social network graph efficiently. 

Finally, ``Predicted Target" in Table~\ref{tab:overview} describes the predicted individual characteristics.  We categorize them into either explicit user characteristics or latent user characteristics. \textit{Explicit User Characteristics} refer to observable user attributes that can be easily collected from a social media platform. The most common explicit user characteristics include demographics  (e.g., age, gender, race etc.), and user preferences (e.g., likes). \textit{Latent User Characteristics} refer to user attributes that cannot be observed easily, and thus need to be assessed by more sophisticated psychometric evaluations. Typical latent user characteristics include personality, human values, depression and addiction. Since typical psychometric tests include dozens or even hundreds of questions (e.g. to assess detailed personality, the IPIP-300 test has 300 questions), it is often difficult to collect the ground truth to assess latent user characteristics at a large scale.

In summary, as shown in table ~\ref{tab:overview}, Twitter is the most common social media platform used in these studies. This is mainly due to  the relative ease of accessing user data on Twitter using its APIs. In terms of the type of user data involved in these studies, text (e.g., tweets or status updates on Facebook) is most common, followed by likes and social networks. Since different types of user data require different analysis techniques (e.g., using graph analytics for social networks and natural language processing for text data), user data type will have significant impact on the machine learning algorithms employed. The size of the datasets varies significantly from as many as 100K people to as few as 383. In general, user attributes declared directly on social media (e.g., topic engagment on Twitter) can be obtained relatively easily at a large scale. In contrast, for user traits/behavior that require sophisticated psychometric evaluation (e.g., depression), their ground truth datasets tend to be smaller. Finally, the predicted user characteristics are quite diverse, ranging from demographics (e.g., age, gender, race, income and occupation), latent user traits (e.g., personality and values), mental disorders (e.g., addiction and depression) and online or real world behaviors (e.g., purchase behavior and cyberbulling).  

In the following, we summarize the typical data analysis methods that are used to infer individual characteristics from social media data.




\section{Inference Methods}
\begin{figure}[t]
\includegraphics[width=0.45\textwidth]{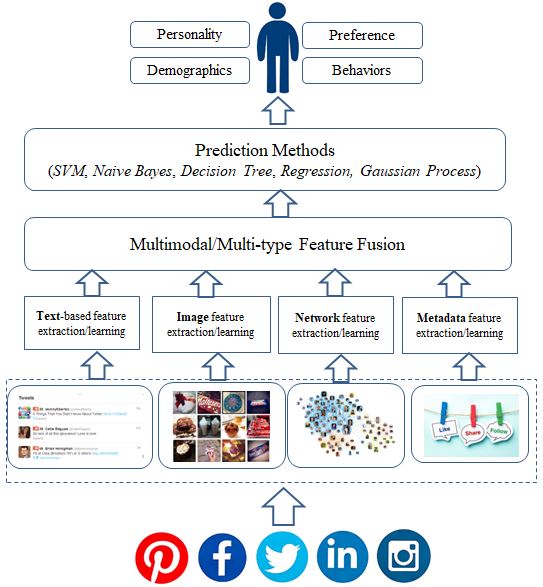}
\caption{A Typical System Architecture}
\label{fig:intro}
\end{figure}

Figure~\ref{fig:intro} shows the typical architecture of such a system. One or more types of user-generated data are extracted from a social media account.  For each type of user data such as text or image, a set of features are extracted (we call this ``single-view feature extraction/learning").  The features from each view are then combined together to form a single user representation (we call this ``multiview feature learning"). Finally, the combined user features are used to predict various human traits and behaviors using typical machine learning methods such as SVM classification or linear/logistic regression.     

There are three main challenges when inferring user traits and behavior from social media data: (1) small labeled training datasets since the ground truth human trait and behavior data are hard to collect at a large scale.  (2) unstructured and high dimensional user data. For example, there could be millions of unigram and bigram features to represent the texts from a user. (3) heterogeneous user data. Frequently we need to combine user data from different modality (e.g., text and images) and types (e.g., structured and unstructured) to paint a complete picture of a user. 

Based on how these challenges are addressed, we categorize the studies in our survey along three dimensions. As shown in ~Table~\ref{tab:MethodOverview}, first, we categorize them based on whether they employ a separate unsupervised feature learning stage to take advantage of a large amount of unsupervised data (2-stage) or they simply use labeled data directly in supervised feature extraction and prediction (1-stage). In general, when the labeled ground truth dataset is small, unsupervised feature learning can boost a system's performance significantly. To address  the second challenge and avoid the ``curse of dimensionality" problem, we categorize these systems based on how they extract, select and learn a small number of features from unstructured data (e.g., text and image). The features can be (a) ``human engineered" where existing  domain/linguistic knowledge is used to construct and select a small number of features or (b) constructed via ``supervised selection", which refers to systems that select relevant features based on their correlations with the ground truth; or (c) constructed via ``unsupervised feature learning", which refers to systems that automatically learn a small number latent features based on unlabeled social media data. In terms of the third challenge, we categorize these studies based on how they combine features from different views together. If they simply concatenate features from different views together, we label them as ``concatenate". If they employ machine learning to fuse information from different views together, we label them as ``fusion". We also list different machine learning algorithms employed for user traits and behavior prediction.   

Since feature extraction, unsupervised feature learning and multiview fusion play important roles in developing such a system, in the following , we describe each topic in detail.

\begin{table*}[ht!]
\small
\centering
    \caption{Summary of Analysis Methods}
    \label{tab:MethodOverview}
    \begin{tabular}{@{}p{4.5cm}p{1cm}p{3.5cm}p{2cm}p{4.5cm}@{}}
      \toprule
       \textbf{Paper} & \textbf{Stage} & \thead{\textbf{Dimension Reduction}}  & \textbf{Fusion}  & \thead{\textbf{Prediction Method}}\\
       \midrule
       \cite{pennacchiotti2011machine} &2-stage &\thead{human engineered \\ unsupervised learning} &concatenation &\thead{Decision Trees} \\\hline 
       \cite{yang2012we} &1-stage&\thead{human engineered}&NA& \thead{SVM} \\\hline
       \cite{de2013predicting} &2-stage& \thead{human engineered \\ unsupervised learning} &concatenation&\thead{SVM}\\\hline 
       \cite{zhang2013predicting} & 2-stage&\thead{human engineered \\ unsupervised learning} &concatenation&\thead{SVM, Naive Bayes\\ Logistic regression }\\\hline 
       \cite{o2013using} & 1-stage& \thead{human engineered} & concatenation&\thead{SVM} \\\hline 
       \cite{kosinski2013private} &2-stage &\thead{unsupervised learning} &NA&\thead{Logistical/Linear regression}\\\hline
       \cite{schwartz2013personality} &2-stage &\thead{human engineered \\ unsupervised learning} & concatenation& \thead{SVM, Linear regression}\\\hline
       \cite{chen2014understanding} &1-stage & \thead{human engineered}&NA &\thead{Logistic regression\\Naive Bayes} \\\hline 
       \cite{gao2014modeling} &1-stage&\thead{supervised selection}&NA&\thead{Lasso regression \\ Least squares regression}\\\hline
       \cite{youyou2015computer} &1-stage &\thead{human engineered\\supervised selection} &NA &\thead{Lasso regression}\\\hline
       \cite{yang2015using} & 1-stage & \thead{human engineered} & NA & \thead{SVM}\\\hline
       \cite{preoctiuc2015analysis} &2-stage &\thead{human engineered \\ unsupervised learning}  &concatenation &\thead{Gaussian Processes}\\\hline
       \cite{lee2015will}&&\thead{human engineered}&concatenation&\thead{SVM, Random Forest\\Naive Bayes, Adaboost\\Logistic regression}\\\hline
       \cite{shen2015study} &1-stage&\thead{human engineered}&concatenation&\thead{Linear Regression}\\\hline
       \cite{bhargava2015unsupervised} &1-stage&\thead{human engineered}&NA& \thead{Hierarchical clustering}\\\hline
       \cite{song2015interest} & 2-stage &\thead{unsupervised learning}&fusion& \thead{SVM, Least squares regression\\ Multi-view multi-task learning}\\\hline
       \cite{hu2016language}&2-stage &\thead{human engineered\\unsupervised learning} &NA&\thead{SVM}\\\hline 
       \cite{song2016volunteerism} & 2-stage &\thead{unsupervised learning} & fusion &\thead{SVM, Decision Trees\\ Random Forest\\Graph-based Learning}\\\hline
       \cite{liu2016analyzing} &1-stage &\thead{human engineered} &NA &\thead{Linear
Regression}\\\hline
       \cite{benton2016learning}&2-stage &\thead{human engineered \\ unsupervised learning}& fusion &\thead{SVM, Multi-task learning}\\\hline
       \cite{ding2017multi} &2-stage &\thead{unsupervised learning} & fusion& \thead{SVM}\\ \hline
       \cite{vedula2017emotional} &1-stage &\thead{human engineered}  & concatenation&\thead{Decision Trees}\\\hline 
         \cite{preoctiuc2017beyond} &2-stage &\thead{human engineered\\ unsupervised learning }&concatenation & \thead{Logistic regression}\\\hline 
         \cite{singh2017toward}&1-stage &\thead{human engineered\\supervised selection} & concatenation&\thead{Bagging classifier}\\ 
      \bottomrule
    \end{tabular}
\end{table*} 

 
\subsection{Basic Feature Extraction}
Basic Feature extraction is often the first step that maps the information in one's social media account into a meaningful and easy to manipulate feature representation. Here, we focus on extracting features from text and images since they are unstructured information and thus more difficult to represent. We will also briefly discuss how to extract features from social networks.

The most commonly used text features are unigrams~\cite{o2013using,preoctiuc2017beyond} and the LIWC (Linguistic Inquiry and Word Count) features.  A unigram is the term frequency computed for each vocabulary word in a text corpus (e.g., a corpus of all the Facebook posts). Sometimes,  unigrams can be weighted based on their informativeness (e.g., based on Inverse Document Frequency or IDF~\cite{benton2016learning}).  Since robust inference often requires repeated word occurrence, low-frequency words are frequently filtered out. In addition to individual words, meaningful phrases can be extracted by keeping only ngrams (e.g., bigrams) with high point-wise mutual information (PMI)~\cite{schwartz2013personality,hu2016language}.   

LIWC features are human engineered features that are constructed based on the psycholinguistic dictionary LIWC. It groups words into psychologically meaningful categories. Empirical results have confirmed that LIWC features are capable of detecting meaning and providing a broader range of social and psychological insights such as feelings, personality, values and motivations. LIWC includes 81 features in five categories such as psychological Processes (e.g., emotional, cognitive, sensory, and social processes), Relativity (e.g., words about time, the past, the future), Personal Concerns (e.g., occupation, financial issues, health), and other dimensions ( e.g., punctuation and swear words). LIWC also includes
writing style features such as word complexity (e.g., words with more than 6 characters). Many of the systems in our survey used LIWC features~\cite{pennacchiotti2011machine,de2013predicting,schwartz2013personality,chen2014understanding,preoctiuc2017beyond,vedula2017emotional,singh2017toward}.  

Sometimes, customized vocabulary is used to extract informative words related to a prediction task. For example, \cite{preoctiuc2017beyond} defined 12,000 political terms in order to select informative unigrams pertaining to politics.

The image data such as profile pictures and photo posts, may contain rich information about individual characteristics. Each image is often represented as a vector of pixels, each pixel is represented by a number (in black and white photos) or three numbers using the RGB color scheme. In addition to the raw image features, meaningful high-level features such as color, facial expressions and postures can be extracted from images ~\cite{liu2016analyzing}, which are then correlated with a user's traits and behavior. Besides general image features, \cite{singh2017toward} extracts demographics and a taxonomy-based object categories (presence of tattoos, graffiti, drug) from Instagram images to facilitate the detection of cyberbullying.


Egocentric social network features are frequently used to characterize the social relations of a user~\cite{de2013predicting,hong2013co,benton2016learning,vedula2017emotional}. An egocentric network is defined as a network containing a single actor (ego), all the actors that an ego is connected to (alters), and all the links between the alters. For each ego, a set of network features are frequently extracted such as network size, betweenness centrality, normalized ego betweenness, cluster coefficient and normalized brokerage. We may also compute measures to assess network homophily such as average age difference between the ego and the alters. 


\subsection{Unsupervised Single View Feature Learning}
Although the ground truth user traits and behavior data are costly to collect at a large scale. it is relatively easy to obtain a large amount of unlabeled user data from social media. The unsupervised feature learning algorithms are used to discover latent features from unlabeled social media data. Here we review typical unsupervised feature learning algorithms. 

\textbf{\em {Singular Value Decomposition (SVD)}} is a mathematical technique that is frequently used for dimension reduction~\cite{svd2000}. Given any $m*n$ matrix A, the algorithm will find matrices $U$, $V$ and $W$ such that $A=UWV^{T}$.  Here $U$ is an orthonormal $m*n$ matrix, $W$ is a diagonal $n*n$ metric and  $V$ is an orthonormal $n*n$ matrix.  Dimensionality reduction is done by computing $R=U*W_r$ where  $W_r$  neglects all but the $r$ largest singular values in the diagonal matrix  $W$. 

\textbf{\em {Principle Component Analysis (PCA)}} is a popular dimensionality reduction mechanism used to eliminate highly correlated variables. PCA can be implemented using SVD.  SVD and PCA have been used to learn a low-dimension representation from a bag-of-word representation of social media posts~\cite{benton2016learning,ding2017multi}, likes~\cite{kosinski2013private,ding2017multi}, and social networks~\cite{benton2016learning}.     

\textbf{{\em Latent Dirichlet Allocation (LDA)}} is a generative graphical model that allows sets of observations to be explained by unobserved latent groups ~\cite{lda2003}.  In natural language processing, LDA is frequently used to learn a set of topics from a large number of documents. The topics are distributions of words that are frequently interpretable. In \cite{schwartz2013personality,ding2017multi}, LDA is employed to learn topics from a user's social media posts. 

\textbf{\em {GloVe}} is an unsupervised learning algorithm originally designed to learn vector representations of words based on aggregated global word-word co-occurrence statistics from a text corpus~\cite{glove2014}. GloVe employs a global log bilinear regression model that combines the advantages of global matrix factorization with that of local context window-based methods. GloVe has been applied to Facebook status updates \cite{preoctiuc2015studying,ding2017multi} and likes \cite{ding2017multi} to learn a dense feature vector for each word/like. To summarize all the words or likes from a user,  we can use a vector aggregation functions such as {\em average}. 

Recently, there is a new generation of neural network-based feature learning methods that employs self-taught learning to automatically derive a feature representation from examples automatically constructed from a large amount of unlabeled social media data. 

\textbf{{\em Autoencoder (AE)} } is a neural network-based feature learning method~\cite{autoencoder2006}. It learns an identity function so that the output is as close to the input as possible. Although an identity function seems a trivial function to learn, by placing additional constraints (e.g,, to make the number of neurons in the hidden layer much smaller than that of the input), we can still force the system to uncover latent structures in the data. 

\textbf{\em {Word Embedding with Word2Vec}} is a neural network-based method originally designed to learn dense vector representations for words~\cite{word2vec2013}. The intuition behind the model is the Distributional Hypothesis, which states words that appear in the same context have similar meanings. There are two models for training a representation of word: continuous bag of word (CBOW) and skipgram (SG) model. CBOW predicts target word from one or more context words, while SG predicts one or more context words from target word. The models are frequently trained using either a hierarchical softmax function (HS) or negative sampling (NS)~\cite{word2vec2013}. To process social media posts, the word2vec model is applied to learn a vector representation of each  word. Then a simple average of all the word vectors by the same user is used to represent all the posts of a user~\cite{benton2016learning,ding2017multi}. In addition, word embeddings can be used to produce word clusters~\cite{preoctiuc2015analysis,preoctiuc2017beyond} and domain lexicons (e.g., depression lexicon) ~\cite{vedula2017emotional}.  

\textbf{\em{Document Embedding with Doc2Vec}} is an extension of Word2Vec, which produces a dense low dimensional feature vector  for a sentence or a paragraph.  There are two Doc2Vec models: Distributed Memory (DM) and Distributed Bag-of-Words (DBOW). Given a sequence of tokens, DM can simultaneously learn a vector representation for each individual word token and a vector for the entire sequence. In DM, each sequence of words (e.g. a paragraph) is mapped to a sequence vector (e.g., paragraph vector) and each word is mapped to a unique word vector.  The paragraph vector and one or more word vectors are aggregated to predict a target word in the context. DBOW learns a global sequence vector to  predict tokens randomly sampled from a sequence. Unlike DM,  DBOW only learns a vector for the entire sequence. It does not learn vectors for individual tokens (e.g., words). Neither does it use a local context window since the words for prediction are randomly sampled from the entire sequence. To characterize an individual's social media posts, a document for each user is created by aggregating every post of the same user.

\subsection{Multi-view Feature Fusion}
To obtain a single, comprehensive and coherent user representation based on all the social media data available, we need to combine user features from different views together. In addition to simply concatenating features extracted from different views, we can also apply  machine learning algorithms to systematically fuse them together. We categorize these fusion methods into two types: (a) unsupervised feature learning that doesn't require any supervised data and (b) supervised multi-task learning. 

There are two types unsupervised learning algorithms for multiview feature fusion : (1) Canonical Correlation Analysis and (b) Deep Canonical Correlation Analysis. 

\textbf{\em {Canonical Correlation Analysis (CCA)}} CCA is a statistical method that explores the relationships between two multivariate sets of variables (vectors)~\cite{hardoon2004canonical}. Given two feature vectors,  CCA tries to find a linear transformation of each feature vector so that they are maximally correlated.  
CCA has been used in ~\cite{sargin2006multimodal,chaudhuri2009multi,kumar2011co,sharma2012generalized,ding2017multi}. 

\textbf{\em {Deep Canonical Correlation Analysis (DCCA)}}
DCCA aims to learn highly correlated deep architectures, which can be a non-linear extension of CCA~\cite{andrew2013deep}. The intuition is to find a maximally correlated representation of two feature vectors by passing them through multiple stacked layers of nonlinear transformation~\cite{andrew2013deep}. Typically, there are three steps to train DCCA: (1) using a denoising autoencoder to pretrain each single view. (2) computing the gradient of the correlation of top-level representation. (3) tuning parameters using back propagation to optimize the total correlation.

The features learned from multiple views are often more informative than those from a single view. Comparing with single-view user feature extraction and learning, multi-view learning achieved significantly better performance in predicting demographics \cite{benton2016learning}, politic leaning ~\cite{benton2016learning} and substance use \cite{ding2017multi}.

\textbf{\em {Multi-task learning} (MTL)} is a supervised learning method to combine multi-view user data together. It tries to jointly train multiple prediction tasks at the same time to exploit the commonalities and differences across tasks.  In \cite{song2016volunteerism}, the authors collected multi-view user data from different platforms (e.g.,  Twitter, Facebook, Linkedin accounts of the same user) and predict volunteerism from two tasks: user-centric analysis and network-centric analysis. Finally, they linearly fused these two components to enhance the final prediction.  




\section{Discussion and Future Directions}
Large-scale social media-based user trait and behavior analysis is an emerging multidisciplinary field with the potential to transform human trait and behavior analysis from controlled small scale experiments to large scale studies of natural human behavior in an open environment. This provides us a new opportunity to explore the interactions of a large number of individual, social and environmental factors simultaneiously,  which would not be possible with traditional methods that use small samples.  The insight gained from these studies could be valuable to help better understand the human minds and decision making process. It will also provide empirical evidence to help public health providers and police makers to improve mental health care and to combat public health threats (e.g., addiction and obese). Due to the privacy concerns on accessing user data on social media and the sensitive nature of the inferred user characteristics, if not careful, there could be significant privacy consequences and ethical implications. So far,  most of the studies in our survey focused primarily on technical contributions. There should be more discussions on ethical considerations when conducting research in this field.    

There are also several promising directions for future research. 
Since individual traits and behavior are highly correlated, building a prediction model that simultaneous infer multiple correlated traits and behavior should yield better performance than predicting each trait/behavior separately. Most existing studies only predict one user attribute/behavior at a time. More research should be conducted to jointly train and predict multiple user attributes together for better performance. 

It is also common for a user to have multiple accounts on different social media platforms. Recently, new technologies have been developed to link different social media accounts of the same user together~\cite{abel2013cross}. With this linked data, it is possible to perform novel cross-platform user trait and behavior analysis such as (1) domain bias analysis that focuses on studying the impact of domain or social media platform on user trait and behavior prediction, (2) domain adaptation that addresses how to adjust prediction models trained on one platform (e.g., Twitter) to predict the traits and behavior on another platform (e.g., Facebook). So far, there is some initial work on domain bias analysis and correction in personality prediction~\cite{kilicc2016analyzing}. More research is needed in order to develop more robust tools for human trait and behavior analysis.  



\clearpage
{\footnotesize
\bibliographystyle{named}
\bibliography{ijcai18}
}

\end{document}